\documentclass[superscriptaddress,msmath,amssymb,aps,prl, twocolumn, epsfig, showpacs, bibliography, lengthcheck]{revtex4-1}

\usepackage{graphicx}
\usepackage{dcolumn}
\usepackage{bm}
\usepackage{hyperref}

\begin{document}
\title{Sub-femtotesla scalar atomic magnetometer using multipass cells}
\author{D. Sheng}
\affiliation{Department of physics, Princeton University, Princeton, New Jersey 08544, USA}
\author{S. Li}
\affiliation{Department of physics, Princeton University, Princeton, New Jersey 08544, USA}
\affiliation{Department of Measurement Technology and Instruments, Zhejiang University of Science and Technology, Hangzhou 310023, China}
\author{N. Dural}
\affiliation{Department of physics, Princeton University, Princeton, New Jersey 08544, USA}
\author{M. V. Romalis}
\affiliation{Department of physics, Princeton University, Princeton, New Jersey 08544, USA}

\date{\today}

\begin{abstract}
Scalar atomic magnetometers have many attractive features but their sensitivity has been relatively poor.  We describe a Rb scalar gradiometer using two multi-pass optical cells. We use a pump-probe measurement scheme to suppress spin-exchange relaxation and two probe pulses to find the spin precession zero crossing times with a resolution of 1 psec. We realize magnetic field sensitivity of 0.54~fT/Hz$^{1/2}$, which improves by an order of magnitude the best scalar magnetometer sensitivity and surpasses the quantum limit set by spin-exchange collisions for a scalar magnetometer with the same measurement volume operating in a continuous regime.

\end{abstract}

\maketitle

Alkali-metal magnetometers can surpass SQUIDs as the most sensitive detectors of magnetic field, reaching sensitivity below 1~fT/Hz$^{1/2}$ \cite{kominis03,Dang}, but only if they are operated near zero magnetic field to eliminate spin relaxation due to spin-exchange collisions \cite{Happer,Allred}. Many magnetometer applications, such as searches for permanent electric dipole moments \cite{WeisEDM}, detection of NMR signals \cite{Ledbetter}, and low-field magnetic resonance imaging \cite{Savukov}, require sensitive magnetic measurements in a finite magnetic field. In addition, scalar magnetometers measuring the Zeeman frequency are unique among magnetic sensors in being insensitive to the direction of the field,  making them particularly suitable for geomagnetic mapping~\cite{Nabighian05} and field measurements in space~\cite{Balogh,Olsen}. The sensitivity of scalar magnetometers has been relatively poor, as summarized recently in \cite{Meyer}. The best directly measured scalar magnetometer sensitivity is equal 7~fT/Hz$^{1/2}$ with a measurement volume of 1.5~cm$^3$~\cite{smullin09}, while estimates of fundamental sensitivity per unit measurement volume for various types of scalar alkali-metal magnetometers range from several fT~cm$^{3/2}$/Hz$^{1/2}$ \cite{budker00,Weis09} to about 1~fT~cm$^{3/2}$/Hz$^{1/2}$ ~\cite{smullin09}. Here we describe a new type of scalar atomic magnetometer using multi-pass vapor cells \cite{Bouchiat,Li12} and operating in a pulsed pump-probe mode \cite{wasilewski10} to achieve magnetic field sensitivity of 0.54~fT/Hz$^{1/2}$ with a measurement volume of 0.66~cm$^3$ in each multi-pass cell. The magnetometer sensitivity  approaches, for the first time, the fundamental limit set by Rb-Rb collisions. We also develop here a quantitative method to analyze significant effects of atomic diffusion on the spectrum of the spin-projection noise in vapor cells with buffer gas using a spin time-correlation function.

The sensitivity of an atomic magnetometer, as any other frequency measurement, is fundamentally limited by spin projection noise and spin relaxation \cite{Huelga97}. For $N$ spin-1/2 atoms with coherence time $T_2$ the sensitivity after a long measurement time $t \gg T_2$ is given by
$\delta B =  \sqrt{2 e/N T_2 t}/\gamma$,
where $\gamma$ is the gyromagnetic ratio. Spin squeezing techniques can reduce this uncertainty by a factor of $\sqrt{e}$, but do not change the scaling with $N$ \cite{Huelga97,Auzinsh04,escher11}. The number of atoms can be increased until collisions between them start to limit $T_2$. Writing $T_2^{-1}=n \sigma \bar{v}$, where $n$ is the density of atoms, $\sigma$ is the spin relaxation cross-section, and $\bar{v}$ is the average collisional velocity, and taking $t=0.5$ sec to calculate the magnetic field spectral noise density $B_n$ in T/Hz$^{1/2}$, we obtain
\begin{equation}
B_n = (2/\gamma) \sqrt{\sigma \bar{v}/V}.
\label{sens}
\end{equation}
Thus the magnetic field spectral noise density per measurement volume $V$ is fundamentally limited by the spin-relaxation cross-section. It also sets the limit on the minimum  energy resolution  per unit bandwidth $\varepsilon=B_n^2 V/2\mu_0$ of atomic magnetometers, which can, in certain cases, approach $\hbar$ \cite{Dang}. In hot alkali-metal vapor magnetometers operating in a finite magnetic field the relaxation is dominated by the spin-exchange cross-section $\sigma_{SE}= 1.9 \times 10^{-14}$ cm$^3$, which gives from Eq.~(\ref{sens}) a limit of about 1.3 fT cm$^{3/2}$/Hz$^{1/2}$ for $\gamma=2\pi\times700$~kHz/G.

However, alkali-metal spin-exchange is a nonlinear process, which can modify the fundamental sensitivity given by Eq.~(\ref{sens}). The magnetic resonance linewidth can be narrowed by optical pumping of atoms into a stretched spin state \cite{appelt98}, but fundamental sensitivity for a scalar magnetometer still remains limited by the spin-exchange cross-section if it is operated in a continuous optical pumping regime \cite{smullin09}. The limit calculated in \cite{smullin09} taking into account $^{87}$Rb nuclear spin $I=3/2$ is 0.51 fT cm$^{3/2}$/Hz$^{1/2}$. On the other hand, if the magnetometer is operated in a pulsed pump-probe regime and takes advantage of spin-squeezing techniques, the sensitivity can be asymptotically limited by the spin-destruction cross-section, which is as low as $\sigma_{SD}= 10^{-18}$ cm$^3$ for K atoms, leading to a potential improvement by 2 orders of magnitude \cite{Vasilakis11}. Thus, it is particularly interesting to study spin projection noise in scalar alkali-metal magnetometers, both because it presents a real limit to their practical sensitivity and because of large improvement possible from spin-squeezing techniques.

\begin{figure}
\includegraphics[width=3.1in]{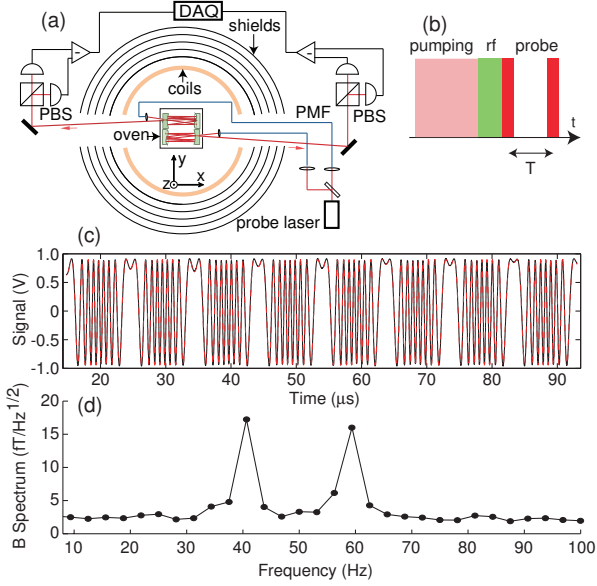}
\caption{\label{fig:setup}(Color online) (a) Experiment setup. PBS: polarized beam splitter, PMF: polarization maintaining fiber, DAQ: data acquisition card. (b) The timing of the pulsed operation. (c) Optical rotation (black line) recorded for one probe pulse at atom density of 0.8$\times$10$^{13}$/cm$^3$ together with a fitted curve (red dash line). (d) Magnetic field noise spectrum obtained in the gradiometer mode in presence of a calibrating gradient magnetic field at 40~Hz.}
\end{figure}

A key parameter for measurements of spin-projection noise is the optical depth on resonance $OD=\sigma_0 n l$, where $\sigma_0$ is probe laser absorption cross-section on resonance and $l$ is the path length of the probe beam through the atomic vapor \cite{Bigelow}. We have developed multi-pass optical cells with mirrors internal to the alkali-metal vapor cell to increase $l$ by two orders of magnitude ~\cite{Li12}. Compared to optical cavities, multi-pass cells have a much larger interaction volume and allow direct recording of large optical rotations.
We use two 42-pass cells placed in the same vapor cell as a gradiometer with a baseline equal to the 1.5 cm distance between the cells, see Fig.~\ref{fig:setup}(a). The cells  have cylindrical mirrors with 10 cm radius of curvature separated by 30 mm. One of the mirrors in each cell has a 2.5 mm diameter hole for entrance and exit of the probe beam focused to a waist diameter of 1.9 mm.  The glass vapor cell contains a drop of enriched $^{87}$Rb and 70 torr N$_2$ gas. A boron-nitride oven is used to heat the vapor cell using AC currents at 600 kHz to 120$^\circ$C, giving an $OD\sim 5000$. The cell is placed in a bias magnetic field of 72.9~mG in the $\hat{z}$ direction generated by an ultra-stable custom current source and is enclosed in a 5-layer magnetic shield.

\begin{figure}
\includegraphics[width=3.1in]{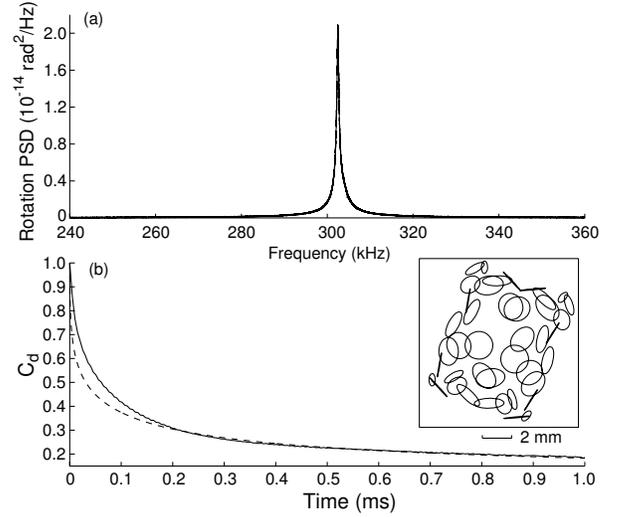}
\caption{\label{fig:sys} (a) Unpolarized atom spin noise spectrum taken at 0.42 G with atom density of 1.2$\times$10$^{13}$/cm$^3$ after subtraction of the photon shot noise background from a spin noise measurement at low magnetic field. (b) Spin correlation function due to the diffusion of the atoms out of the probe beam. Solid line (dashed line) is the experimental (calculated) result with atom density of 1.2$\times$10$^{13}$/cm$^3$. The inset shows the calculated beam pattern at the center of the cavity.}
\end{figure}

We measure the atom density $n$ from the transverse relaxation $T_{2}$ at low polarization, which is dominated by spin-exchange collisions with a known cross-section~\cite{Li12}. The  number of atoms participating in the measurement at any given time $N=n V_b$ is determined from the area of Faraday rotation power spectral density for unpolarized atoms~\cite{Shah10}. We make measurements of the noise peak at two different magnetic fields and take their difference to remove the background dominated by photon shot noise.  Fig.~\ref{fig:sys}(a) shows one example of unpolarized power spectral density obtained using this method, which gives $V_b= 0.35(2)$~cm$^3$ for each cell.

While diffusion does not affect the area under the spin noise peak, it causes the lineshape of the noise spectrum to deviate from a simple Lorentzian. To analyze it quantitatively, we consider the time autocovariance function of the Faraday rotation signal $\phi(t)$, which is given by the Fourier transform of the power spectrum. One can show that
\begin{eqnarray}
&&\left\langle \phi(t) \phi (t+\tau)\right\rangle =\sum_{i}\left(\frac{cr_{e}f D_i}{(2I+1)\int I(\mathbf{r})dydz}\right)^{2} \times\nonumber\\
&&n\left\langle F_{i}^{2}\right\rangle\int I(\mathbf{r}_{1})G(\mathbf{r}_{1}-\mathbf{r}_{2},\tau) I(\mathbf{r}_{2})d^{3}\mathbf{r}_{1}d^{3}\mathbf{r}_{2},
\end{eqnarray}
where the sum is taken over the two alkali-metal hyperfine states, $F_a=I+1/2$ and $F_b=I-1/2$, and $\left\langle F_{i}^{2}\right\rangle=F_i(F_i+1)(2F_i+1)/6(2I+1)$. The dispersion factor $D_i=1/(\nu_i-\nu)$ for far detuning of the probe frequency $\nu$ from the hyperfine resonances $\nu_i$. Here $I(\mathbf{r})$ is the total probe laser intensity at position $\mathbf{r}$, including all beam passes inside the cavity,  and $G(r,\tau)$ is the Green's function for  spin evolution with a diffusion coefficient $D$ and a transverse relaxation time $T_2$,
$G(r,\tau)=e^{-r^2/4D\tau-\tau/T_2}/(4{\pi}D\tau)^{3/2}$. The intensity profile of the probe laser in the cell is determined by measuring the input Gaussian beam size and calculating the astigmatic Gaussian beam propagation in the multi-pass cell~\cite{Kasyutich}. An example of the calculated intensity profile in the middle of the cell is shown in the inset of Fig.~2. The effective number of atoms participating in the measurement is defined as the number of atoms that would generate the same spin noise area $\left\langle \phi(t)^2\right\rangle$ if interrogated with a uniform probe intensity.  We obtain a generalization of a result given in Ref.~\cite{Shah10} that works for laser beams with varying focusing and overlap,
\begin{equation}
N=n l^2\frac{\left(\int{I(\mathbf{r})}dydz\right)^2}{\int{I(\mathbf{r})^2}dV},
\end{equation}
where $l$ is the total probe laser path in the multi-pass cell. Based on calculated intensity profile we obtain $V_b=0.36$~cm$^3$, in good agreement with direct experimental measurements. In Fig.~2 we compare the diffusion component of the calculated spin time-correlation function  $C_d(\tau)=\left\langle \phi(t)\phi (t+\tau )\right\rangle e^{t/T_{2}}/\left\langle \phi(t)^2\right\rangle$ with the experimental measurement obtained  by taking the Fourier transform of the spin noise peak after centering it at zero frequency and correcting for the transverse spin relaxation time  $C_{d}(\tau)=C(\tau)e^{\tau/T_{2}}$. They agree well except at early times due to deviations from a perfect Gaussian of tightly focused beams within the cell, indicating that Green's function method can quantitatively describe the lineshape of the spin noise spectrum in the presence of diffusion with multiple overlapping laser beams.

Figure~\ref{fig:setup}(b) shows the timing for magnetic field measurements.  For optical pumping pulse, which lasts 14 msec, we use two circular polarized beams on resonance with the $D1$ transitions from both ground hyperfine states. Then we apply a $\pi/2$ rf pulse lasting 3 periods of the Zeeman resonance frequency. We apply the first probe light pulse shortly after the rf excitation and the second probe pulse with a delay time $T$ from the first one. The probe laser is tuned to 794.780~nm and the power of the light exiting from each multi-pass cell is about 0.5~mW.  We turn on and off the probe light slowly compared with the Larmor period using an AOM to suppress transient spin excitation. The pump-probe cycle is repeated every 16.6 msec, synchronized with  60~Hz to reduce its influence.

Figure~\ref{fig:setup}(c) shows a typical record of the optical rotation signal during one of the probe pulses. We fit the data using the equation~\cite{Li12}:
\begin{equation}
V=A\sin\left(2\phi\left(1-\frac{t-t_c}{T_2}\right)\sin(\omega(t-t_c))+\psi\right)+B.
\label{eq:fit}
\end{equation}
We find the time of zero crossings $t_{c1}$,$t_{c2}$ of the first and second pulse and calculate $T_c=t_{c2}-t_{c1}$ which gives a measure of the magnetic field $B=2\pi m/\gamma T_c$, where $m$ is the number of spin precession cycles between measured zero crossings.  If the measurements are repeated with overall duty cycle $d$, then the magnetic field sensitivity per Hz$^{1/2}$ is given by:
\begin{equation}
B_n = B \delta T_c \sqrt{2/d T_c },
\label{eq:sens}
\end{equation}
where $\delta T_c$ is the standard deviation of repeated measurements of $T_c$.

The two multi-pass cells work as a gradiometer to measure $\partial{B_z}/\partial{y}$ with a noise level which is $\sqrt{2}$ larger than given by Eq.~(\ref{eq:sens}). While a scalar magnetometer does not require calibration, we check its response by applying a known source of magnetic field gradient.  Fig.~\ref{fig:setup}(c) shows Fourier transform of repeated magnetic field measurements in the presence of a gradient $\partial{B_z}/\partial{y}$ with rms amplitude of 21.6~fT/cm oscillating at 40 Hz. It introduces a 33~fT magnetic field difference between the centers of the two cells. For this measurement the atomic density is 1.4$\times$10$^{13}$/cm$^3$, with the probe pulse length of four Larmor periods, the separation between two probe pulses $T$ of 823~$\mu$s, and the cycle period is 5~ms. The integration of 40~Hz peak in the spectrum gives the rms signal of 32.4~fT and confirms the sensitivity of the gradiometer.

\begin{figure}
\includegraphics[width=3.1in]{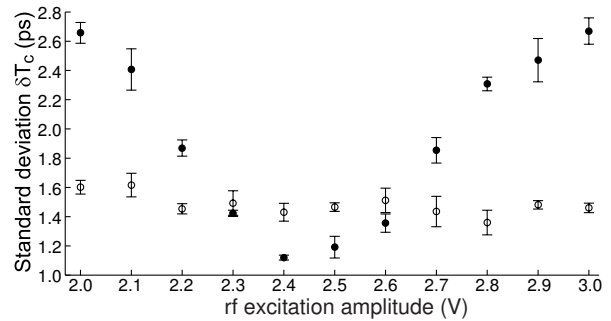}
\caption{\label{fig:bak}  The dependence of the standard deviation $\delta T_c$ on the rf pulse amplitude. Empty (solid) circles denote the back-action evasion with stroboscopic modulation (normal modulation) of probe light. Here the probe pulse length and $T$ are equal to four Larmor periods.}
\end{figure}

The limiting fundamental noise sources include spin projection noise (SPN) and photon shot noise (PSN), while technical sources include magnetic shield noise and time jitter of the data acquisition. One important feature of our arrangement is back-action evasion of quantum fluctuations of the probe beam circular polarization due to zero spin polarization of atoms in the $\hat{z}$ direction following the $\pi/2$ pulse.  Fig.~\ref{fig:bak} shows the dependence of the noise on the rf excitation amplitude when it deviates from the $\pi/2$ amplitude. We compare it to the noise in $T_c$ when using a stroboscopic probe modulation back-action evasion scheme~\cite{Vasilakis11}, where the probe beam is modulated at twice the Larmor frequency with a duty cycle 20\% in the probe pulse period. The results confirm that the magnetometer works in back-action-free regime. The magnetic shield gradient noise is due by thermal Johnson currents and is calculated based on known electrical conductivity of the shields, giving 0.40(5)~fT/Hz$^{1/2}$~\cite{lee08}. The time jitter noise is determined by recording the signals from the same multi-pass cell with two acquisition channels and ranges from 0.3 to 0.5 ps depending on the length of the probe pulse.

\begin{figure}
\includegraphics[width=3.1in]{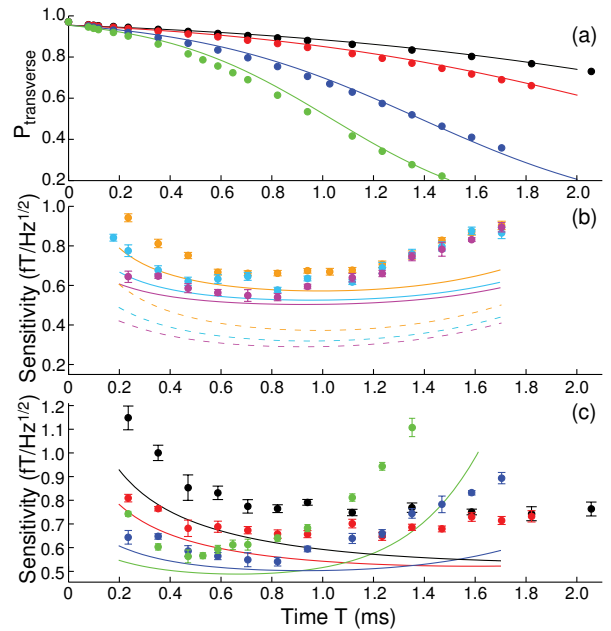}
\caption{\label{fig:sig}(color) (a) Relaxation of transverse polarization $P_t$ (points) for $n=$1.9 (green), 1.4 (blue), 0.8 (red) and 0.6 (black) $\times$10$^{13}$/cm$^3$ together with theoretical prediction (lines).  (b) The points are the experimental results of magnetic field sensitivity at $n$=1.4$\times$10$^{13}$/cm$^3$, with probe pulse length of four (orange), eight (cyan) and twelve (purple) Larmor periods. Solid (dashed) lines are the theoretically calculated sensitivity based on measured parameters with (without) magnetic shield noise and time jitter noise. (c) Experiment results (points) of magnetic field sensitivity with probe length of twelve Larmor periods at different atom densities, with the same color notation as plot~(a), together with theoretical predictions (lines).}
\end{figure}

 Fig.~\ref{fig:sig} shows the magnetic field sensitivity for a range of parameters. The experimental duty cycle is limited to 10\% by available pump laser power, but we use $d=1$ in Eq.~(\ref{eq:sens}) to find the fundamental sensitivity from the measured uncertainty $\delta T_c$.   Fig.~\ref{fig:sig}(a) shows the nonlinear relaxation of transverse spin polarization due to spin exchange at four different densities, from which we find that the initial transverse polarization is equal to 0.96(1). We plot the sensitivity as a function of the probe pulse length $t_p$ in Fig.~\ref{fig:sig}(b). The variance in $T_c$ due to PSN and data acquisition noise decreases as $1/t_p$ and the variance due to SPN also decreases because atom diffusion effectively involves more atoms into the measurement. The effective number of atoms $N_m$ participating in the measurement after a pulse time $t_p$ can be found using the diffusion correlation function $N_m=nV_b/\left[2/t_p\int_0^{t_p}(1-t/t_p)C_d(t)dt\right]$. For the longest pulse length of 230 $\mu$sec we obtain $V_m=1.9V_b$, corresponding to an effective interaction volume of 0.66~cm$^3$. We also show theoretical estimate of the sensitivity including ASN, PSN, magnetic gradient noise and time jitter noise in Fig.~\ref{fig:sig}(b) with solid lines and only ASN and PSN with broken lines.  Figure~\ref{fig:sig}(c) shows similar results at other densities. When the atom density increases, the optimal $T$ decreases because of faster spin relaxation, indicating that the magnetometer works in Rb collision-limited regime.   For the longest probe pulse length and atom density of 1.4$\times$10$^{13}$/cm$^3$, the experimental data shows a best sensitivity of 0.54~fT/Hz$^{1/2}$, which is 10\% above the predicted value. In the absence of magnetic shield noise the fundamental sensitivity is projected to be $0.3$~fT/Hz$^{1/2}$, dominated by ASN.

In conclusion, we described a scalar magnetometer based on multipass atomic vapor cells. It uses a pulsed mode with a high initial polarization and reaches the spin-exchange collision limited regime where the sensitivity is largely independent of atom density. The best sensitivity obtained is 0.54~fT/Hz$^{1/2}$ with an effective interaction volume of 0.66~cm$^3$, which is an  order of magnitude improvement over the previous best sensitivity for a scalar magnetometer and exceeds the quantum-limited sensitivity of a scalar magnetometer operating in a continuous mode.   We also developed a quantitative method for analyzing the effect of diffusion on quantum spin noise using spin time-correlation function. By relying on precision timing measurements with very wide dynamic range and fractional field sensitivity of $7\times 10^{-11}/$Hz$^{1/2}$ this magnetometer  opens the possibility of fundamentally new applications, for example unshielded detection of magnetoencephalography signals~\cite{Xia06}. The sensitivity per unit volume can be further improved in this system by reducing the decay of the spin time-correlation function due to atomic diffusion, which will allow suppression of ASN due to spin-squeezing between two probe pulses. The spin correlation decay is dominated by a few tightly focused beam spots in the multi-pass cell and can be reduced by modifying multi-pass cell parameters to avoid tight beam focusing. This work was supported by DARPA.

\bibliography{magnetometer}

\end{document}